\documentclass{article}

\usepackage{amsfonts}
\usepackage{amssymb}
\usepackage{amsmath}
\usepackage{latexsym}
\usepackage{amsthm}
\usepackage{eucal}
\usepackage{eufrak}
\usepackage[dvips]{graphicx}
\usepackage{psfrag}

\author{A. Cerri$^{1,2}$\quad P. Frosini$^{1,2}$\quad C.
Landi$^{1,3,}$\footnote{Corresponding author. E-mail Address:
\texttt{clandi@unimore.it}}\\\vspace{-0.3cm}
\small{$^{1}$ ARCES, Universit\`a di Bologna,}\\
\small{via Toffano $2/2$, I-$40135$ Bologna,
Italia}\\\vspace{-0.3cm} \small{$^{2}$Dipartimento di Matematica,
Universit\`a di
Bologna,}\\
\small{P.zza di Porta S. Donato 5, I-$40126$ Bologna,
Italia}\\\vspace{-0.3cm}
\small{$^3$DISMI, Universit\`a di Modena e Reggio Emilia,}\\
\small{via Amendola $2$, Pad. Morselli, I-$42100$ Reggio Emilia,
Italia}}

\title{Stability in multidimensional Size Theory}

\newcommand{\R}{\mathbb R}

\newcommand{\x}{\vec x}
\newcommand{\y}{\vec y}

\newcommand{\fr}{\vec \varphi}
\newcommand{\eps}{\varepsilon}
\newcommand{\M}{{\mathcal M}}

\newtheorem{prop}{Proposition}
\newtheorem{definition}{Definition}
\newtheorem{lemma}{Lemma}
\newtheorem{theorem}{Theorem}
\newtheorem{oss}{Remark}
\newtheorem{cor}{Corollary}

\begin{document}

\maketitle

\begin{abstract} This paper proves that in Size Theory the
comparison of multidimensional size functions can be reduced to
the $1$-dimensional case by a suitable change of variables.
Indeed, we show that a foliation in half-planes can be given, such
that the restriction of a multidimensional size function to each
of these half-planes turns out to be a classical size function in
two scalar variables. This leads to the definition of a new
distance between multidimensional size functions, and to the proof
of their stability with respect to that distance.
\end{abstract}
\vspace{1cm}

\section*{Introduction} Shape comparison is probably one of the
most challenging issues in Computer Vision and Pattern
Recognition. In recent years many papers have been devoted to this
subject and new mathematical techniques have been developed to
deal with this problem.
%As an example among many, the
%importance of making available new approaches for comparing shapes
%of proteins has been focused by the research group of Edelsbrunner
%at Duke University (cf....).
In the early 90's, Size Theory was proposed as a
geometrical/topological approach to shape comparison. The main
idea is to translate the comparison of two datasets (e.g.
3D-models, images or sounds) into the comparison of two suitable
topological spaces $\mathcal{M}$ and $\mathcal{N}$, endowed with
two continuous functions $\vec{\varphi}:\mathcal{M}\to \R^k$,
$\vec{\psi}:\mathcal{N}\to \R^k$. These functions are called
\emph{$k$-dimensional measuring functions} and are chosen
according to the application. In other words, they can be seen as
descriptors of the properties considered relevant for the
comparison. In \cite{FrMu99} the definition of the natural
pseudo-distance $d$ between the pairs
$(\mathcal{M},\vec{\varphi})$, $(\mathcal{N},\vec{\psi})$ was
introduced, setting
$d\left((\mathcal{M},\vec{\varphi}),(\mathcal{N},\vec{\psi})\right)$
equal to the infimum of the change of the measuring function,
induced by composition with all the homeomorfisms from
$\mathcal{M}$ to $\mathcal{N}$. Unfortunately, the study of $d$ is
quite difficult, even for $k=1$, although strong properties can be
proved in this case (cf. \cite{DoFr04,DoFr07}). Size Theory
tackles this problem by introducing some mathematical tools that
allow us to easily obtain lower bounds for $d$, such as \emph{size
homotopy groups} and \emph{size functions}  (cf.
\cite{FrMu99} and \cite{DoFr04bis}). The idea is to study the pairs
$(\mathcal{M}\langle\fr\preceq
\x\,\rangle,\mathcal{M}\langle\fr\preceq \y\,\rangle)$, where
$\mathcal{M}\langle\fr\preceq \vec t\ \rangle$ is defined by
setting $\mathcal{M}\langle\fr\preceq\vec t\
\rangle=\{P\in\mathcal{M}:\varphi_i(P)\leq t_i,i=1,\ldots,k\}$ for
$\vec{t}=(t_1,\dots,t_k)\in \R^k$. The $k$-th size homotopy group
$\pi_{k}(\x,\y)$ describes the
 non-trivial equivalence classes of $k$-dimensional loops in $\mathcal{M}\langle\fr\preceq
\x\,\rangle$ that remain homotopically non-trivial also in
$\mathcal{M}\langle\fr\preceq \y\,\rangle$. Size functions count
the number of connected components in
$\mathcal{M}\langle\fr\preceq \y\,\rangle$ that meet
$\mathcal{M}\langle\fr\preceq \x\,\rangle$. It turns out that
$\pi_{0}(\x,\y)$ is a set whose cardinality is equal to the value
taken by the size function at $(\x,\y)$. From the homological
point of view, an analogous approach, named \emph{Size Functor},
has been developed in \cite{CaFePo01} for $1$-dimensional
measuring functions.

More recently, similar ideas have independently led Edelsbrunner
et al. to the definition of \emph{Persistent Homology} (cf.
\cite{EdLeZo00,EdLeZo02}), and Allili et al. to the definition of
the \emph{Morse Homology Descriptor} (cf. \cite{AlCoZi04}).

From the beginning of the 90's, size functions have been studied
and applied in the case of $1$-dimensional measuring functions
(cf., e.g.,
\cite{DiFrPa04,Fr90,Fr91,Fr96,FrLa97,KaMiMr04,UrVe97,VeUr93,VeUr96,VeUrFrFe93}).
The multidimensional case presented more severe difficulties,
since a concise, complete and stable description of
multidimensional size functions was not available before this
work.

In \cite{CaZo06}, Carlsson and Zomorodian  examine the
completeness problem by studying  Multidimensional Persistent
Homology. In that paper, it is claimed that multidimensional
persistence has an essentially different character from its
$1$-dimensional version. Indeed, their approach does not seem to
lead to a concise, complete and stable descriptor in the
multidimensional case, whereas it does in classical Persistent
Homology (see \cite{CoEdHa05}).

The first result of this paper is the proof that in Size Theory
the comparison of multidimensional size functions can be reduced
to the $1$-dimensional case by a suitable change of variables
(Theorem \ref{reduction}). The key idea is to show that a
foliation in half-planes can be given, such that the restriction
of a multidimensional size function to these half-planes turns out
to be a classical size function in two scalar variables. Our
approach implies that each size function, with respect to a
$k$-dimensional measuring function, can be completely and
compactly described by a parameterized family of discrete
descriptors (Remark \ref{Descriptor}). This follows from the
results proved in \cite{FrLa01} about the representation of
classical size functions by means of formal series of points and
lines, applied to each plane in our foliation. An important
consequence is that we can easily prove the stability of this new
descriptor (and hence of the corresponding $k$-dimensional size
function) also in higher dimensions (Proposition \ref{stability}),
by using a recent result of stability proved for $1$-dimensional
size functions (cf. \cite{dAFrLa,dAFrLabis}) and analogous to the
result obtained in \cite{CoEdHa05} for Persistence Homology. As a
final contribution, we show that a matching distance between size
functions, with respect to measuring functions taking values in
$\R^k$, can easily be introduced (Definition \ref{multidimatch}).
This matching distance provides a lower bound for the natural
pseudo-distance, also in the multidimensional case (Theorem
\ref{lowbound}). All these results open the way to the use of
Multidimensional Size Theory in real applications.\\
\textbf{Outline.} In Section \ref{Definitions} we give the
definition of $k$-dimensional size function. In Section
\ref{Reduction} the foliation we use is presented, and the
reduction to the $1$-dimensional case is proved. Section
\ref{Lowb} shows the stability of our computational method,
implying a lower bound for the natural pseudo-distance.
Additionally, a new distance between multidimensional size
functions is introduced. In Section \ref{Example} the
effectiveness of the multidimensional approach is tested on an
example. Section \ref{links} examines some links between
multidimensional size functions and the concept of
\emph{vineyard}, recently introduced in \cite{CSEdMo06}. Section
\ref{Conclusions} concludes the paper, presenting some discussion
and future work.

\section{Definition of k-dimensional size function}\label{Definitions}
For the present paper, $\mathcal{M}$, $\mathcal{N}$ denote two
non-empty compact and locally connected Hausdorff spaces.
\vspace{0.4cm}
\newline
In Multidimensional Size Theory \cite{FrMu99}, any pair
$(\mathcal{M},\fr)$, where
$\fr=(\varphi_1,\dots,\varphi_k):\mathcal{M}\rightarrow\R^k$ is a
continuous function, is called a \emph{size pair}. The function
$\fr$ is called a \emph{$k$-dimensional measuring function}. The
following relations $\preceq$ and $\prec$ are defined in $\R^k$:
for $\x=(x_1,\dots,x_k)$ and $\y=(y_1,\dots,y_k)$, we say
$\x\preceq\y$ (resp. $\x\prec\y$) if and only if $x_i\leq\ y_i$
(resp. $x_i<y_i$) for every index $i=1,\dots,k$. Moreover, $\R^k$
is endowed with the usual $\max$-norm:
$\|(x_1,x_2,\dots,x_k)\|_{\infty}=\max_{1\le i\le k}|x_i|$. In
this framework, if $\M$ and $\mathcal{N}$ are homeomorphic, the
size pairs $(\mathcal{M},\fr)$ and $(\mathcal{N},\vec{\psi})$ can
be compared by means of the \emph{natural pseudo-distance} $d$,
defined as
$$
d((\mathcal{M},\fr),(\mathcal{N},\vec{\psi}))=\inf_{f}\max_{P\in\mathcal{M}}\|\fr(P)-\vec{\psi}(f(P))\|_{\infty}\
,
$$
where $f$ varies among all the homeomorphisms between
$\mathcal{M}$ and $\mathcal{N}$. The term pseudo-distance means
that $d$ can vanish even if the size pairs do not coincide. Here,
and in what follows, $\R^k\times\R^k$ and $\R^{2k}$ are
identified.

Now we introduce the $k$-dimensional analogue of size function for
a size pair $(\mathcal{M},\fr)$. We shall use the following
notations: $\Delta^+$ will be the open set
$\{(\x,\y)\in\R^k\times\R^k:\x\prec \y\}$, while
$\Delta=\partial\Delta^+$. For every $k$-tuple
$\x=(x_1,\dots,x_k)\in\R^k$, let $\mathcal{M}\langle\fr\preceq
\x\,\rangle$  be the set $\{P\in\mathcal{M}:\varphi_i(P)\leq x_i,\
i=1,\dots,k\}$.

\begin{definition}
For every $k$-tuple $\y=(y_1,\dots,y_k)\in\R^k$, we shall say that
 two points $P,Q\in \mathcal{M}$ are
$\langle\fr\preceq \y\,\rangle$-\emph{connected} if and only if a
connected subset of $\mathcal{M}\langle\fr\preceq \y\,\rangle$
exists, containing $P$ and $Q$.
\end{definition}

\begin{definition}
We shall call \emph{($k$-dimensional) size function} associated
with the size pair $(\mathcal{M},\fr)$ the function
$\ell_{(\mathcal{M},\fr)}:\Delta^+\rightarrow\mathbb{N}$, defined
by setting $\ell_{(\mathcal{M},\fr)}(\x,\y)$ equal to the number
of equivalence classes in which the set
$\mathcal{M}\langle\fr\preceq \x\,\rangle$ is divided by the
$\langle\fr\preceq \y\,\rangle$-connectedness relation.
\end{definition}

\begin{oss}\label{altdef}
In other words, $\ell_{(\mathcal{M},\fr)}(\x,\y)$ counts the
connected components in $\mathcal{M}\langle\fr\preceq \y\,\rangle$
containing at least one point of $\mathcal{M}\langle\fr\preceq
\x\,\rangle$.
\end{oss}

\section{Reduction to the 1-dimensional case}\label{Reduction}

In this section, we will show that there exists a parameterized
family of half-planes in $\R^k\times\R^k$ such that the
restriction of $\ell_{(\mathcal{M},\fr)}$ to each of these planes
can be seen as a particular 1-dimensional size function.
\begin{definition}
For every unit vector $\vec{l}=(l_1,\ldots,l_k)$ of $\mathbb{R}^k$
such that $l_i>0$ for every $i=1,\dots,k$, and for every vector
$\vec{b}=(b_1,\ldots,b_k)$ of $\mathbb{R}^k$ such that
$\sum_{i=1}^k b_i=0$, we shall say that the pair
$(\vec{l},\vec{b})$ is \emph{admissible}. We shall denote the set
of all admissible pairs in $\R^k\times\R^k$ by $Adm_k$. Given an
admissible pair $(\vec{l},\vec{b})$, we define the half-plane
$\pi_{(\vec{l},\vec{b})}$ of $\R^k\times\R^k$ by the following
parametric equations:
$$
\left\{%
\begin{array}{ll}
    \vec x=s\vec l + \vec b\\
    \vec y=t\vec l + \vec b\\
\end{array}%
\right.
$$
for $s,t\in \R$, with $s<t$.
\end{definition}

\begin{prop}
For every $(\vec{x},\vec{y})\in \Delta^+$ there exists one and
only one admissible pair $(\vec{l},\vec{b})$ such that
$(\vec{x},\vec{y})\in \pi_{(\vec{l},\vec{b})}$.
\end{prop}
\begin{proof}
The claim immediately follows by taking, for $i=1,\dots,k$,
$$l_i=\frac{y_i-x_i}{\sqrt{\sum_{j=1}^k(y_j-x_j)^2}}\ ,\quad b_i=\frac{x_i\sum_{j=1}^k y_j-y_i\sum_{j=1}^k x_j}{\sum_{j=1}^k
(y_j-x_j)}.$$ Then, $\vec x=s\vec l + \vec b$, $\vec y=t\vec l +
\vec b$, with
\begin{eqnarray*}
s=\frac{\sum_{j=1}^k x_j}{\sum_{j=1}^k l_j}=\frac{\sum_{j=1}^k
x_j\sqrt{\sum_{j=1}^k(y_j-x_j)^2}}{\sum_{j=1}^k(y_j-x_j)}\\
t=\frac{\sum_{j=1}^k y_j}{\sum_{j=1}^k l_j}=\frac{\sum_{j=1}^k
y_j\sqrt{\sum_{j=1}^k(y_j-x_j)^2}}{\sum_{j=1}^k(y_j-x_j)}.\\
\end{eqnarray*}
\end{proof}

Now we can prove the reduction to the $1$-dimensional case.

\begin{theorem}\label{reduction}
Let $(\vec{l},\vec{b})$ be an admissible pair, and $F_{(\vec
l,\vec b)}^{\fr}:\mathcal{M}\rightarrow\R$ be defined by setting
$$
F_{(\vec l,\vec
b)}^{\fr}(P)=\max_{i=1,\dots,k}\left\{\frac{\varphi_i(P)-b_i}{l_i}\right\}\
.
$$
Then, for every $(\vec x,\vec y)=(s\vec l+\vec b,t\vec l + \vec
b)\in\pi_{(\vec{l},\vec{b})}$ the following equality holds:
$$
\ell_{(\mathcal{M},\fr)}(\vec x,\vec
y)=\ell_{(\mathcal{M},F_{(\vec l,\vec b)}^{\fr})}(s,t)\ .
$$
\end{theorem}
\begin{proof}
For every $\vec{x}=(x_1,\dots,x_k)\in\R^k$, with $x_i=sl_i+b_i,\
i=1,\dots,k$, it holds that
$\mathcal{M}\langle\fr\preceq\vec{x}\,\rangle=\mathcal{M}\langle
F_{(\vec l,\vec b)}^{\fr}\leq s\rangle$. This is true because
\begin{eqnarray*}\mathcal{M}\langle\fr\preceq\vec{x}\,\rangle&=&\{P\in\mathcal{M}:
\varphi_i(P)\leq x_i,\ i=1,\dots,k\}=\\
&=&\{P\in\mathcal{M}:
\varphi_i(P)\leq sl_i+b_i,\ i=1,\dots,k\}=\\
&=&\left\{P\in\mathcal{M}:\frac{\varphi_i(P)-b_i}{l_i}\leq s, \
i=1,\dots,k\right\}= \mathcal{M}\langle F_{(\vec l,\vec
b)}^{\fr}\leq s\rangle.
\end{eqnarray*}
Analogously, for every $\vec{y}=(y_1,\dots,y_k)\in\R^k$, with
$y_i=tl_i+b_i,\ i=1,\dots,k$, it holds that
$\mathcal{M}\langle\fr\preceq\vec{y}\,\rangle=\mathcal{M}\langle
F_{(\vec l,\vec b)}^{\fr}\leq t\rangle$. Therefore Remark
\ref{altdef} implies the claim.
\end{proof}

In the following, we shall use the symbol $F_{(\vec l,\vec
b)}^{\fr}$ in the sense of Theorem \ref{reduction}.

\begin{oss}\label{Descriptor}
Theorem \ref{reduction} allows us to represent each
multidimensional size function as a parameterized family of formal
series of points and lines, on the basis of the description
introduced in \cite{FrLa01} for the $1$-dimensional case. Indeed,
we can associate a formal series $\sigma_{(\vec l,\vec b)}$ with
each admissible pair $(\vec l,\vec b)$, with $\sigma_{(\vec l,\vec
b)}$ describing the $1$-dimensional size function
$\ell_{(\mathcal{M},F_{(\vec l,\vec b)}^{\fr})}$. The family
$\left\{\sigma_{(\vec l,\vec b)}:(\vec l,\vec b)\in Adm_k\right\}$
is a new complete descriptor for $\ell_{(\mathcal{M},\vec
\varphi)}$, in the sense that two multidimensional size functions
coincide if and only if the corresponding parameterized families
of formal series coincide.
\end{oss}

\section{Lower bounds for the k-dimensional natural
pseudo-distance}\label{Lowb}

In \cite{dAFrLa,dAFrLabis}, it has been shown that $1$-dimensional
size functions can be compared by means of a distance, called
\emph{matching distance}. This distance is based on the
observation that each $1$-dimensional size function is the sum of
characteristic functions of triangles. The matching distance is
computed by finding an optimal matching between the sets of
triangles describing two size functions. For a formal definition
we refer to \cite{dAFrLa,dAFrLabis} (see also \cite{CoEdHa05} for
the analogue of the matching distance in Persistent Homology). In
the sequel, we shall denote by
$d_{match}(\ell_{(\mathcal{M},F_{(\vec l,\vec
b)}^{\fr})},\ell_{(\mathcal{N},F_{(\vec l,\vec b)}^{\vec\psi})})$
the matching distance between the $1$-dimensional size functions
$\ell_{(\mathcal{M},F_{(\vec l,\vec b)}^{\fr})}$ and
$\ell_{(\mathcal{N},F_{(\vec l,\vec b)}^{\vec\psi})}$.

The following result is an immediate consequence of Theorem
\ref{reduction} and Remark \ref{Descriptor}.

\begin{cor}
Let us consider the size pairs $(\mathcal{M},\fr)$,
$(\mathcal{N},\vec\psi)$. Then, the identity
$\ell_{(\mathcal{M},\fr)}\equiv\ell_{(\mathcal{N},\vec{\psi})}$
holds if and only if $d_{match}(\ell_{(\mathcal{M},F_{(\vec l,\vec
b)}^{\fr})},\ell_{(\mathcal{N},F_{(\vec l,\vec
b)}^{\vec\psi})})=0$, for every admissible pair
$(\vec{l},\vec{b})$.
\end{cor}

The next result proves that small enough changes in $\fr$ with respect to
the $\max$-norm induce small changes of
$\ell_{(\mathcal{M},F_{(\vec l,\vec b)}^{\fr})}$ with respect to
the matching distance.

\begin{prop}\label{stability}
If $(\mathcal{M},\fr)$, $(\mathcal{M},\vec{\chi})$ are size pairs
and $\max_{P\in\mathcal{M}}
\|\fr(P)-\vec\chi(P)\|_{\infty}\leq\epsilon$, then for each
admissible pair $(\vec{l},\vec{b})$, it holds that
$$
d_{match}(\ell_{(\mathcal{M},F_{(\vec l,\vec
b)}^{\fr})},\ell_{(\mathcal{M},F_{(\vec l,\vec
b)}^{\vec\chi})})\leq\frac{\epsilon}{\min_{i=1,\dots,k}l_i}.
$$
\end{prop}
\begin{proof}
 From the Matching
Stability Theorem $25$ in \cite{dAFrLa} (see also
\cite{dAFrLabis})  we obtain that
\begin{eqnarray*}
d_{match}(\ell_{(\mathcal{M},F_{(\vec l,\vec
b)}^{\fr})},\ell_{(\mathcal{M},F_{(\vec l,\vec
b)}^{\vec\chi})})&\leq&\max_{P\in\mathcal{M}} |F_{(\vec l,\vec
b)}^{\fr}(P)-F_{(\vec l,\vec
b)}^{\vec\chi}(P)|.
\end{eqnarray*}
Let us now fix $P\in\M$. Then, denoting by $\hat \iota$ the
index for which $\max_i
\frac{\varphi_i(P)-b_i}{l_i}$ is attained,  by the definition of $F_{(\vec l,\vec
b)}^{\fr}$ and $F_{(\vec l,\vec b)}^{\vec\chi}$ we have that
\begin{eqnarray*}
\ & &  F_{(\vec l,\vec
b)}^{\fr}(P)-F_{(\vec l,\vec
b)}^{\vec\chi}(P)= \max_i
\frac{\varphi_i(P)-b_i}{l_i}-
\max_i \frac{\chi_i(P)-b_i}{l_i}= \\
& = & \frac{\varphi_{\hat \iota}(P)-b_{\hat {\iota}}}{l_{\hat \iota}}-
\max_i \frac{\chi_i(P)-b_i}{l_i}\le  \frac{\varphi_{\hat \iota}(P)-b_{\hat
    {i}}}{l_{\hat \iota}}-\frac{\chi_{\hat \iota}(P)-b_{\hat
    {\iota}}}{l_{\hat \iota}}=\\
& = & \frac{\varphi_{\hat \iota}(P)-\chi_{\hat \iota}(P)}{l_{\hat \iota}}\le
\frac{\|\fr(P)-\vec\chi(P)\|_{\infty}}{\min_{i=1,\dots,k}l_i}. 
\end{eqnarray*}
In the same way,  we obtain 
$F_{(\vec l,\vec
b)}^{\vec\chi}(P)-F_{(\vec l,\vec
b)}^{\fr}(P)  \le \frac{\|\fr(P)-\vec\chi(P)\|_{\infty}}{\min_{i=1,\dots,k}l_i}.$
Therefore, if $\max_{P\in\mathcal{M}}
\|\fr(P)-\vec\chi(P)\|_{\infty}\leq\epsilon$,
$$ \max_{P\in\mathcal{M}}\left|F_{(\vec l,\vec
b)}^{\fr}(P)-F_{(\vec l,\vec
b)}^{\vec\chi}(P)  \right|\leq  \max_{P\in\mathcal{M}}
\frac{\|\fr(P)-\vec\chi(P)\|_{\infty}}{\min_{i=1,\dots,k}l_i}\le \frac{\epsilon}{\min_{i=1,\dots,k}l_i}.$$
\end{proof}

\begin{oss}\label{lStability}
Analogously, it is easy to show that small enough changes in $(\vec l,
\vec b)$ with respect to the $\max$-norm induce small changes of
$\ell_{(\mathcal{M},F_{(\vec l,\vec b)}^{\fr})}$ with respect to
the matching distance.
\end{oss}

Proposition \ref{stability} and Remark \ref{lStability} prove the
stability of our computational approach.

Now we are able to prove our next result, showing that a lower
bound exists for the multidimensional natural pseudo-distance.

\begin{theorem}\label{lowbound}
Let $(\mathcal{M},\fr)$ and $(\mathcal{N},\vec{\psi})$ be two size
pairs, with $\mathcal{M}$, $\mathcal{N}$ homeomorphic. Setting
$d((\mathcal{M},\fr),(\mathcal{N},\vec{\psi}))=\inf_{f}\max_{P\in\mathcal{M}}\|\fr(P)-\vec{\psi}(f(P))\|_{\infty}$,
where $f$ varies among all the homeomorphisms between
$\mathcal{M}$ and $\mathcal{N}$, it holds that
$$
\sup_{(\vec l,\vec b)\in Adm_k}\min_{i=1,\dots,k}l_i\cdot
d_{match}(\ell_{(\mathcal{M},F_{(\vec l,\vec
b)}^{\fr})},\ell_{(\mathcal{N},F_{(\vec l,\vec
b)}^{\vec\psi})})\leq
d((\mathcal{M},\fr),(\mathcal{N},\vec{\psi})).
$$
\end{theorem}
\begin{proof}
For any homeomorphism $f$ between $\mathcal{M}$ and $\mathcal{N}$,
it holds that $\ell_{(\mathcal{N},F_{(\vec l,\vec
b)}^{\vec\psi})}\equiv\ell_{(\mathcal{M},F_{(\vec l,\vec
b)}^{\vec\psi}\circ f)}$. Moreover, by applying Proposition
\ref{stability} with
$\epsilon=\max_{P\in\mathcal{M}}\|\fr(P)-\vec\psi(f(P))\|_{\infty}$
and $\vec\chi=\vec\psi\circ f$, and observing that
$F^{\vec\psi}_{(\vec l,\vec b)}\circ f\equiv F^{\vec\psi\circ
f}_{(\vec l,\vec b)}$, we have
$$
\min_{i=1,\dots,k}l_i\cdot d_{match}(\ell_{(\mathcal{M},F_{(\vec
l,\vec b)}^{\fr})},\ell_{(\mathcal{N},F_{(\vec l,\vec
b)}^{\vec\psi})})
\leq\max_{P\in\mathcal{M}}\|\fr(P)-\vec\psi(f(P))\|_{\infty}$$ for
every admissible $(\vec{l},\vec{b})$. Since this is true for each
homeomorphism $f$ between $\mathcal{M}$ and $\mathcal{N}$, the
claim immediately follows.
\end{proof}

\begin{oss}
We observe that the left side of the inequality in Theorem
\ref{lowbound} defines a distance between multidimensional size
functions associated with homeomorphic spaces. When the spaces are not assumed to be homeomorphic, it still verifies all the properties of a
distance, except for the fact that it may take the value $+\infty$. In other words, it defines
an extended distance.
\end{oss}

\begin{definition}\label{multidimatch}
Let $(\mathcal{M},\fr)$ and $(\mathcal{N},\vec{\psi})$ be two size
pairs. We shall call \emph{multidimensional matching distance} the
extended distance defined by setting
$$
D_{match}(\ell_{(\mathcal{M},\fr)},\ell_{(\mathcal{N},\vec\psi)})=\sup_{(\vec
l,\vec b)\in Adm_k}\min_{i=1,\dots,k}l_i\cdot
d_{match}(\ell_{(\mathcal{M},F_{(\vec l,\vec
b)}^{\fr})},\ell_{(\mathcal{N},F_{(\vec l,\vec b)}^{\vec\psi})}).
$$
\end{definition}

\begin{oss}
\label{A} If we choose a non-empty subset $A\subseteq Adm_k$ and
we substitute $\sup_{(\vec l,\vec b)\in Adm_k}$ with $\sup_{(\vec
l,\vec b)\in A}$ in Definition \ref{multidimatch}, we obtain an (extended)
pseudo-distance between multidimensional size functions. If $A$ is
finite, this pseudo-distance appears to be particularly suitable
for applications, from a computational point of view.
\end{oss}

\section{An example}\label{Example}

\begin{figure}[h]
\begin{center}
\psfrag{R2}{\!\!\!{\small$\!\!t=\mathbf{\sqrt{2}}$}}
\psfrag{R1}{\!\!{\small$t=\mathbf{1}$}} \psfrag{M}{$\mathcal{M}$}
\psfrag{N}{$\mathcal{N}$} \psfrag{C}{$\ell_{(\mathcal{M},F_{(\vec
l,\vec b)}^{\fr})}$} \psfrag{D}{\ $\ell_{(\mathcal{N},F_{(\vec
l,\vec b)}^{\vec\psi})}$}
\begin{tabular}{cc}
\includegraphics[height=3.4cm]{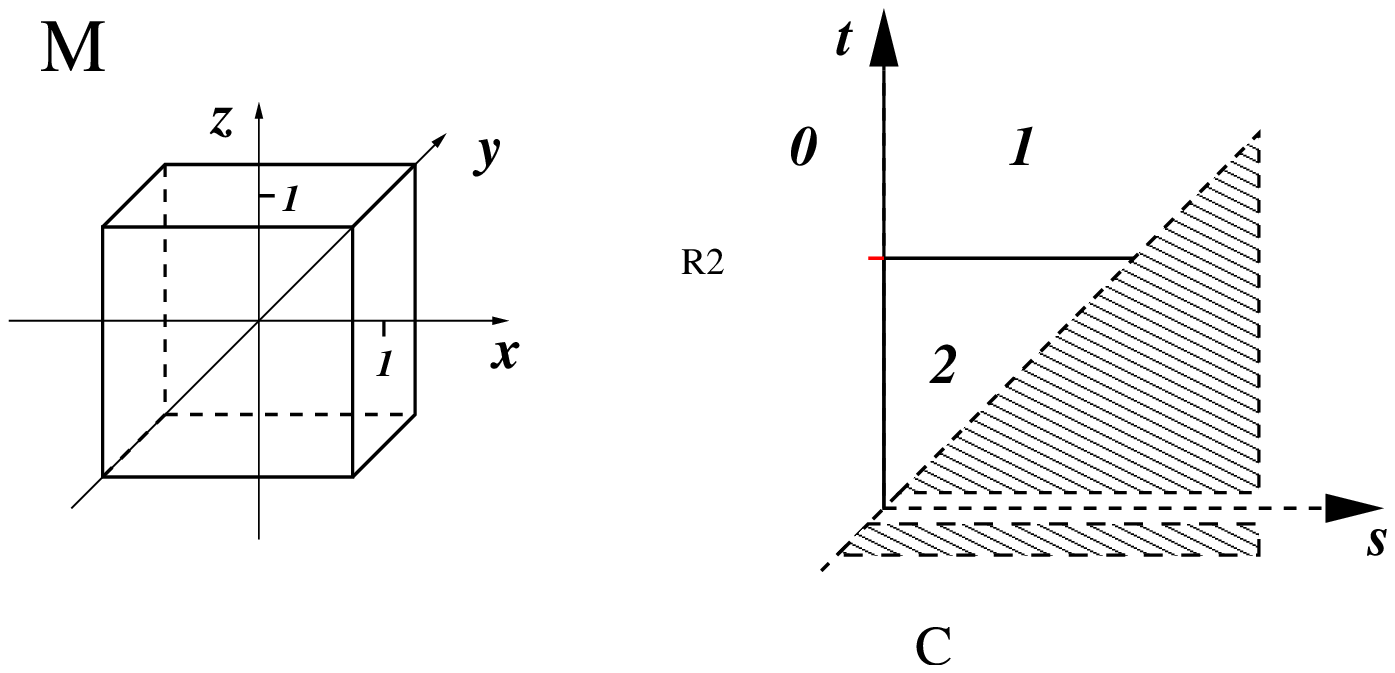}&
\includegraphics[height=3.4cm]{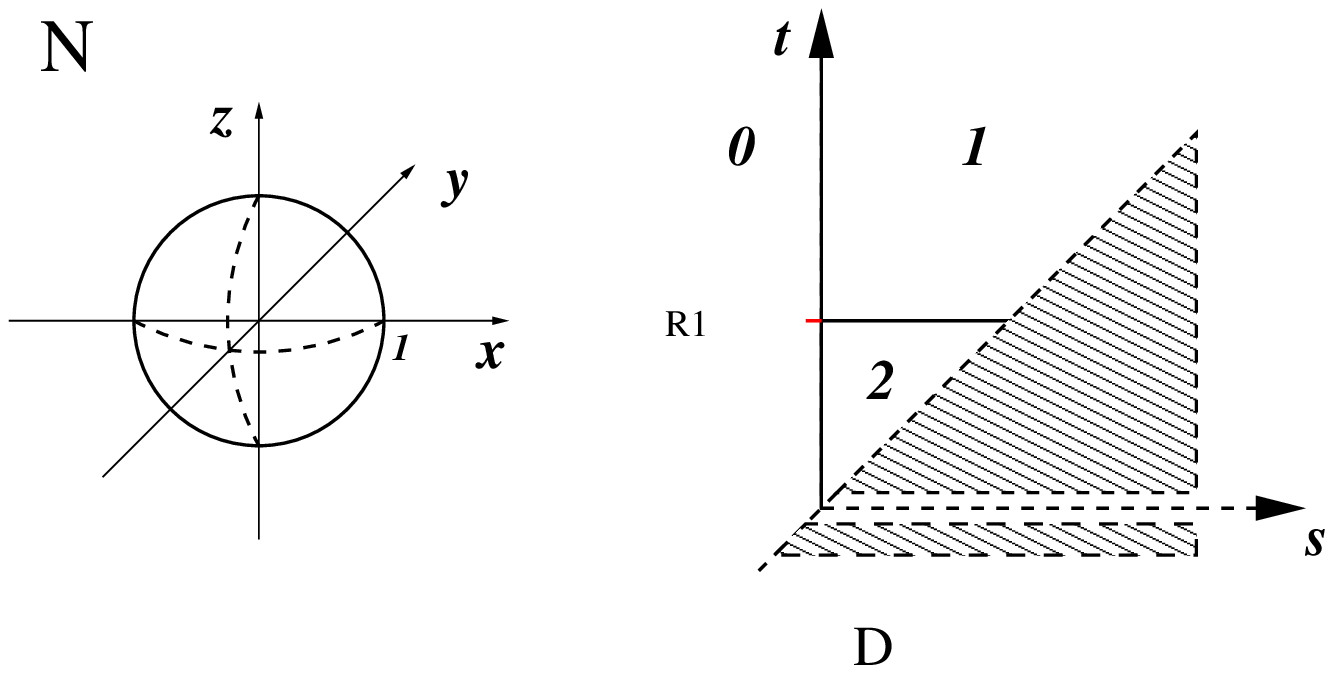}
\end{tabular}
\caption{The topological spaces $\mathcal{M}$ and $\mathcal{N}$
and the size functions $\ell_{(\mathcal{M},F_{(\vec l,\vec
b)}^{\fr})},\ell_{(\mathcal{N},F_{(\vec l,\vec b)}^{\vec\psi})}$
associated with the half-plane $\pi_{(\vec l,\vec b)}$, for $\vec
l=(\frac{\sqrt{2}}{2},\frac{\sqrt{2}}{2})$ and $\vec b=(0,0)$.}
\label{figura:FigArt}
\end{center}
\end{figure}

In $\R^3$ consider the set
$\mathcal{Q}=[-1,1]\times[-1,1]\times[-1,1]$ and the sphere
$\mathcal{S}$ of equation $x^2+y^2+z^2=1$. Let also
$\vec\Phi=(\Phi_1,\Phi_2):\R^3\rightarrow\R^2$ be the continuous
function, defined as $\vec\Phi(x,y,z)=(|x|,|z|)$. In this setting,
consider the size pairs $(\mathcal{M},\fr)$ and
$(\mathcal{N},\vec{\psi})$ where
$\mathcal{M}=\partial\mathcal{Q}$, $\mathcal{N}=\mathcal{S}$, and
$\fr$ and $\vec\psi$ are respectively the restrictions of
$\vec\Phi$ to $\mathcal{M}$ and $\mathcal{N}$. In order to compare
the size functions $\ell_{(\mathcal{M},\fr)}$ and
$\ell_{(\mathcal{N},\vec\psi)}$, we are interested in studying the
foliation in half-planes $\pi_{(\vec l,\vec b)}$, where $\vec
l=(\cos\theta,\sin\theta)$ with $\theta\in(0,\frac{\pi}{2})$, and
$\vec b=(a,-a)$ with $a\in\R$. Any such half-plane is represented
by
$$
\left\{%
\begin{array}{ll}
    x_1=s\cos\theta + a\\
    x_2=s\sin\theta - a\\
    y_1=t\cos\theta + a\\
    y_2=t\sin\theta - a\\
\end{array}%
\right.\ ,
$$
with $s,t\in\R$, $s<t$. Figure \ref{figura:FigArt} shows the size
functions $\ell_{(\mathcal{M},F_{(\vec l,\vec b)}^{\fr})}$ and
$\ell_{(\mathcal{N},F_{(\vec l,\vec b)}^{\vec\psi})}$, for
$\theta=\frac{\pi}{4}$ and $a=0$, i.e. $\vec
l=\left(\frac{\sqrt{2}}{2},\frac{\sqrt{2}}{2}\right)$ and $\vec
b=(0,0)$. With this choice, we obtain that $F_{(\vec l,\vec
b)}^{\fr}=\sqrt{2}\max\{\varphi_1,\varphi_2\}=\sqrt{2}\max\{|x|,|z|\}$
and $F_{(\vec l,\vec
b)}^{\vec\psi}=\sqrt{2}\max\{\psi_1,\psi_2\}=\sqrt{2}\max\{|x|,|z|\}$.
Therefore, Theorem \ref{reduction} implies that, for every
$(x_1,x_2,y_1,y_2)\in\pi_{(\vec l,\vec b)}$
\begin{eqnarray*}
\ell_{(\mathcal{M},\fr)}(x_1,x_2,y_1,y_2)=\ell_{(\mathcal{M},\fr)}\left(\frac{s}{\sqrt{2}},\frac{s}{\sqrt{2}},\frac{t}{\sqrt{2}},\frac{t}{\sqrt{2}}\right)=\ell_{(\mathcal{M},F_{(\vec l,\vec b)}^{\fr})}(s,t)\\
\ell_{(\mathcal{N},\vec\psi)}(x_1,x_2,y_1,y_2)=\ell_{(\mathcal{N},\vec\psi)}\left(\frac{s}{\sqrt{2}},\frac{s}{\sqrt{2}},\frac{t}{\sqrt{2}},\frac{t}{\sqrt{2}}\right)=\ell_{(\mathcal{N},F_{(\vec
l,\vec b)}^{\vec\psi})}(s,t)\ .
\end{eqnarray*}
In this case, by Theorem \ref{lowbound} and Remark \ref{A}
(applied for $A$ containing just the admissible pair that we have
chosen), a lower bound for the natural pseudo-distance
$d((\mathcal{M},\fr),(\mathcal{N},\vec{\psi}))$ is given by
$$
\frac{\sqrt{2}}{2}d_{match}(\ell_{(\mathcal{M},F_{(\vec l,\vec
b)}^{\fr})},\ell_{(\mathcal{N},F_{(\vec l,\vec b)}^{\vec\psi})})
=\frac{\sqrt{2}}{2}(\sqrt{2}-1)=1-\frac{\sqrt{2}}{2}.
$$
Indeed, the matching distance
$d_{match}(\ell_{(\mathcal{M},F_{(\vec l,\vec
b)}^{\fr})},\ell_{(\mathcal{N},F_{(\vec l,\vec b)}^{\vec\psi})})$
is equal to the cost of moving the point of coordinates
$(0,\sqrt{2})$ onto the point of coordinates $(0,1)$, computed
with respect to the $\max$-norm. The points $(0,\sqrt{2})$ and
$(0,1)$ are representative of the characteristic triangles of the
size functions $\ell_{(\mathcal{M},F_{(\vec l,\vec b)}^{\fr})}$
and $\ell_{(\mathcal{N},F_{(\vec l,\vec b)}^{\vec\psi})}$,
respectively.
\\We conclude by observing that
$\ell_{(\mathcal{M},\varphi_1)}\equiv\ell_{(\mathcal{N},\psi_1)}$
and
$\ell_{(\mathcal{M},\varphi_2)}\equiv\ell_{(\mathcal{N},\psi_2)}$.
In other words, the multidimensional size functions, with respect
to $\fr,\vec\psi$, are able to discriminate the cube and the
sphere, while both the $1$-dimensional size functions, with
respect to $\varphi_1,\varphi_2$ and $\psi_1,\psi_2$, cannot do
that. The higher discriminatory power of multidimensional size
functions motivates their definition and use.

\section{Links between dimension $\bf 0$ vineyards and multidimensional size
functions}\label{links}

In a recent paper \cite{CSEdMo06}, Cohen-Steiner et al. have
introduced the concept of {\em vineyard}, that is a $1$-parameter
family of persistence diagrams associated with the homotopy $f_t$,
interpolating between $f_0$ and $f_1$. These authors assume that
the topological space is homeomorphic to the body of a simplicial
complex, and that the measuring functions are \emph{tame}. We
shall do the same in this section. We recall that dimension $p$
persistence diagrams are a concise representation of the function
$rank\,H^{x, y}_p$, where $H^{x, y}_p$ denotes the dimension $p$
persistent homology group computed at point $(x,y)$ (cf.
\cite{CSEdMo06}). Therefore, the information described by
vineyards is equivalent to the knowledge of the function
$rank\,H^{x, y}_p$, computed with respect to the function $f_t$.
We are interested in the case $p=0$. Since, by definition, for
$x<y$, $rank\,H^{x, y}_0$ coincides with the value taken by the
size function $\ell_{(\M,f_{t})}(x,y)$, it follows that, for
$x<y$, dimension $0$ vineyards contain the same information as the
$1$-parameter family of size functions
$\{\ell_{(\M,f_{t})}\}_{t\in[0,1]}$. Anyway, another interesting
link exists between dimension $0$ vineyards and multidimensional
size functions. This link is expressed by the following theorem.
In order to prove it, we need the next two lemmas. The former
states that the relation of $\langle\fr\preceq \vec
y\,\rangle$-connectedness passes to the limit.

\begin{lemma}\label{ycont}
Assume that $(\M,\vec \varphi)$ is a size pair and $\vec y
=(y_1,\ldots,y_k)\in \R^k$. If, for every $\eps> 0$, $P$ and $Q$
are $\langle\fr\preceq
(y_1+\eps,\ldots,y_k+\eps)\,\rangle$-connected in $\M$, then they
are also $\langle\fr\preceq \vec y\,\rangle$-connected.
\end{lemma}
\begin{proof}
For every positive integer number $n$, let $K_{n}$ be the
connected component of
$\M\langle\fr\preceq(y_1+\frac{1}{n},\dots,y_k+\frac{1}{n})\rangle$
containing $P$ and $Q$. Since connected components are closed sets
and  $\M$ is compact, each $K_{n}$ is compact. The set $\bigcap_n
K_{n}$ is the intersection of a family of connected compact
Hausdorff subspaces with the property that $K_{n+1}\subseteq
K_{n}$ for every $n$, and hence it is connected (cf. Theorem
$28.2$ in \cite{Wi70} p. $203$). Moreover, $\bigcap_n K_{n}$ is a
subset of $\M\langle\fr\preceq\vec y\rangle$ and contains both $P$
and $Q$. Therefore, $P$ and $Q$ are $\langle\fr\preceq \vec
y\,\rangle$-connected.
\end{proof}

The following lemma allows us to study the behavior of
multidimensional size functions near $\Delta$ (where they have not
been defined because of instability problems when the measuring
functions are not assumed to be tame).

\begin{lemma}\label{limit}
Let $(\M,\fr)$ be a size pair. If $\vec x\preceq \vec y$ then
$\lim_{\eps\to 0^+}\ell_{(\M,\fr)}\left(\vec
x,(y_1+\eps,\ldots,y_k+\eps)\right)$ is equal to the number
$L(\vec x,\vec y)$ of equivalence classes of $\M\langle \fr\preceq
\vec x\rangle$ quotiented  with respect to the $\langle\fr\preceq
\vec y\,\rangle$-connectedness relation.
\end{lemma}
Note that, for $\vec x\prec\vec y$, $L(\vec x,\vec y)$ simply
coincides with $\ell_{(\M,\fr)}\left(\vec x,\vec y\right)$.

\begin{proof}[Proof of Lemma \ref{limit}]
First of all we observe that the function
$\ell_{(\M,\fr)}\left(\vec x,(y_1+\eps,\ldots,y_k+\eps)\right)$ is
nonincreasing in the variable $\eps$, and hence the value
$\lim_{\eps\to 0^+}\ell_{(\M,\fr)}\left(\vec
x,(y_1+\eps,\ldots,y_k+\eps)\right)$ is defined. The statement of
the theorem is trivial if $\lim_{\eps\to
0^+}\ell_{(\M,\fr)}\left(\vec
x,(y_1+\eps,\ldots,y_k+\eps)\right)=+\infty$, since, for every
$\eps>0$, the inequality $\ell_{(\M,\fr)}\left(\vec
x,(y_1+\eps,\ldots,y_k+\eps)\right)\le L(\vec x,\vec y)$ holds by
definition, and hence the equality $L(\vec x,\vec y)=+\infty$
immediately follows. Let us now assume that $\lim_{\eps\to
0^+}\ell_{(\M,\fr)}\left(\vec
x,(y_1+\eps,\ldots,y_k+\eps)\right)=r<+\infty$. In this case a
finite set $\mathcal{P}=\{P_1,\ldots,P_r\}$ of points in
$\M\langle \fr\preceq \vec x\rangle$ exists such that, for every
small enough $\eps>0$, every $P\in \M\langle \fr\preceq \vec
x\rangle$ is $\langle\vec
\varphi\preceq(y_1+\eps,\ldots,y_k+\eps)\rangle$-connected to a
point $P_j\in \mathcal{P}$ in $\M$. Furthermore, for $i\neq j$ the
points $P_i,P_j\in \mathcal{P}$ are not $\langle\vec
\varphi\preceq(y_1+\eps,\ldots,y_k+\eps)\rangle$-connected and
hence not $\langle\vec \varphi\preceq\y\,\rangle$-connected
either. From Lemma \ref{ycont} it follows that every $P\in
\M\langle \fr\preceq \vec x\rangle$ is $\langle\vec
\varphi\preceq\vec y)\rangle$-connected to a point $P_j\in
\mathcal{P}$ in $\M$. Therefore $L(\vec x,\vec y)=r=\lim_{\eps\to
0^+}\ell_{(\M,\fr)}\left(\vec
x,(y_1+\eps,\ldots,y_k+\eps)\right)$.
\end{proof}

\begin{theorem}
For $t\in I=[0,1]$, consider the family of size pairs $(\M,f_t)$
where $f_t$ is a homotopy between $f_0:\M\rightarrow\R$ and
$f_1:\M\rightarrow\R$. Define $\vec \chi:\M\times
I\rightarrow\R^3$ by $\vec\chi(P,t)=(f_t(P),t,-t)$. Then, for
every $ \bar t\in I$ and $ \bar x, \bar y\in \R$ with $ \bar x\le
\bar y$, it holds that
$$
rank\,H_0^{ \bar x, \bar y}(\bar t)= \lim_{\eps\to
0^+}\ell_{(\M\times I,\vec \chi)}( \bar x, \bar t,- \bar t,  \bar
y+\eps,  \bar t+\eps, - \bar t +\eps),
$$
where $H^{\bar x, \bar y}_0(\bar t)$ denotes the dimension $0$
persistent homology group computed at point $(\bar x,\bar y)$ with
respect to $f_{\bar t}$.
\end{theorem}
\begin{proof}
We know that $rank\,H_0^{ \bar x, \bar y}(\bar t)$ is equal to the
number of equivalence classes of $\M\langle f_{\bar t} \le \bar
x\rangle$ quotiented with respect to the $\langle f_{\bar t} \le
\bar y\rangle$-connectedness relation. On the other hand, Lemma
\ref{limit} states that $\lim_{\eps\to 0^+}\ell_{(\M\times I,\vec
\chi)}( \bar x, \bar t,-\bar t, \bar y + \eps, \bar t+\eps, - \bar
t +\eps)$ is equal to the number of equivalence classes of
$\M\times I\langle \vec\chi\preceq( \bar x, \bar t, - \bar
t)\rangle$ quotiented, with respect to the $\langle
\vec\chi\preceq( \bar y, \bar t, - \bar t)\rangle$-connectedness
relation. By definition of $\vec \chi$, this last number equals
the number of equivalence classes of $\M\langle f_{\bar t} \le
\bar x\rangle$ quotiented, with respect to the $\langle f_{\bar t}
\le \bar y\rangle$-connectedness relation. This concludes our
proof.
\end{proof}

However, although these two links exist, the concept of
multidimensional size function has quite different purposes than
that of vineyard. First of all, vineyards are based on a
$1$-parameter parallel foliation of $\R^3$, while the study of
multidimensional size functions depends on a $(2k-2)$-parameter
non-parallel foliation of $\Delta^+\subseteq\R^k\times\R^k$. In
fact, multidimensional size functions are associated with
$k$-dimensional measuring functions, instead of with a homotopy
between $1$-dimensional measuring functions. Furthermore,
\cite{CSEdMo06} does not aim to identify distances for the
comparison of vineyards, while we are interested in quantitative
methods for comparing multidimensional size functions.

\section{Conclusions and future work}\label{Conclusions}

In this paper we have proved that the theory of multidimensional
size functions can be reduced to the $1$-dimensional case by a
suitable change of variables. This equivalence implies that
multidimensional size functions are stable, with respect to the
new distance $D_{match}$, and this allows us to use them in
concrete applications by exploiting the existing computational
techniques.

Many theoretical problems deserve further investigation, among
them we list a few here.
\begin{itemize}
    \item \textbf{Choice of the foliation.} Other foliations,
    different from the one we propose are possible. In general,
    we can choose a family $\Gamma$ of continuous curves
    $\vec\gamma_{\vec \alpha}:\R\rightarrow\R^k$ such that (i) for $s<t$,
    $\vec\gamma_{\vec \alpha}(s)\prec\vec\gamma_{\vec \alpha}(t)$, (ii) for every
    $(\vec x,\vec y)\in\Delta^+$ there is one and only one
    $\vec\gamma_{\vec \alpha}\in\Gamma$ through $\vec x$, $\vec y$ and (iii)
    the curve $\gamma_{\vec \alpha}$ depends continuously on the parameter
    $\vec \alpha$ (this last hypothesis is important in computation for
    stability reasons). It would be interesting to study the dependence
    of our results on the choice of the foliation.
    \item \textbf{Extension to the algebraic context.} We think
    that the main results obtained in this paper for multidimensional size
    functions can be straightforwardly extended to the ranks of size homotopy groups
    and persistent homology groups.
\end{itemize}

\subsection*{Acknowledgements}
Work performed within the activity of ARCES ``E. De Castro'',
University of Bologna, under the auspices of INdAM-GNSAGA, of the
University of Bologna, funds for selected research topics,
and of the University of Modena and Reggio Emilia.\\
This paper is dedicated to the memory of Marco Gori.

\end{document}